\documentstyle[11pt,newpasp,epsfig,twoside]{article}
\begin{document}
\title{Verifying the use of supernovae as probes of the cosmic expansion}

\author{Richard~Ellis$^{1,2}$ and Mark~Sullivan$^{2,1}$}

\affil{$^1$105-24 Astronomy, Caltech, Pasadena, CA 91125, USA}
\affil{$^2$Institute of Astronomy, Madingley Road, Cambridge, CB3 0HA, UK}

\begin{abstract}
  We present preliminary results of a follow-up survey which aims
  to characterise in detail those galaxies which hosted Type Ia 
  supernovae found by the Supernova Cosmology Project. Our survey 
  has two components: Hubble Space Telescope imaging with STIS and 
  Keck spectroscopy with ESI, the goal being to classify each host
  galaxy into one of three broad morphological/spectral classes and 
  hence to investigate the dependence of supernovae properties on host 
  galaxy type over a large range in redshift. Of particular interest 
  is the supernova Hubble diagram characterised by host galaxy class 
  which suggests that most of the scatter arises from those occurring 
  in late-type irregulars. Supernovae hosted by (presumed dust-free) 
  E/S0 galaxies closely follow the adopted SCP cosmological model.
  Although larger datasets are required, we cannot yet find any
  significant difference in the light curves of distant supernovae 
  hosted in different galaxy types.
\end{abstract}

\section{Introduction}

The discovery and controlled systematic study of faint, distant type
Ia supernovae (SNe\,Ia) has transformed the prospects for constraining
the time derivative of the cosmic expansion rate.  Though
well-measured local SNe\,Ia have a `raw' observed dispersion in their
maximum luminosity of $\simeq0.35\,\rm{mags}$, empirical correlations
between the peak SNe luminosity and light-curve width enable this scatter
to be reduced (Phillips 1993; Hamuy et al. 1996b; Riess et al.
1996; Perlmutter et al. 1999). By using such light curve shape corrections 
with independent samples of SNe\,Ia, two Hubble diagrams have been
constructed by the High-Redshift Supernovae Search Team (Riess et al.
1998) and the Supernovae Cosmology Project (SCP, Perlmutter et al.
1999, hereafter P99). These data strongly exclude the
hitherto popular Einstein de Sitter cosmology ($\Omega=1$,
$\Lambda=0$). In combination with the results of recent microwave
background measurements (de Bernardis et al. 2000; Jaffe et al. 2000)
which indicate a spatially flat inflationary universe, the SNe\,Ia results
suggest a significant non-zero cosmological constant ($\Omega=0.28$,
$\Lambda=0.72$).

Conclusions of this importance require excellent supporting evidence.
In particular, it is appropriate to question the homogeneity,
environmental trends and evolutionary behaviour of the SNe found at all
redshifts. Systematic differences between high and low redshift samples
might change the derived cosmological parameters without destroying the
small dispersion seen in the SNe\,Ia Hubble diagrams. Evolutionary
differences between low and high-redshift SNe\,Ia might arise via the
progenitor composition (H\"{o}flich 1999), a differential dust
extinction with greater amounts of dust in high-redshift environments,
either in the host galaxy (Totani \& Kobayashi 1999) or `grey' dust in
the IGM (Aguirre 1999), or a dependence of the SNe properties on host
galaxy environments. Addressing this last possibility forms the 
basis of this present study.

Though SNe\,Ia can occur in all types of galaxies, disk and spheroidal
stellar populations should sample different ranges of metallicities and
dust content, and thus we might expect that SNe\,Ia progenitor
composition and the light curve properties could be affected
accordingly. By studying SNe grouped according to the mean underlying
stellar population, we can directly investigate such effects. 

Local studies of the environmental effects of local SNe\,Ia have been
conducted by Hamuy et al. (1996a, 2000), Branch et al. (1996) and Riess
et al. (1999). In various low-redshift samples, they find the
distribution of light curve decline rates correlates with the
morphological type of the host galaxy, with ellipticals hosting faster
decline rate (dimmer) SNe, whereas spiral galaxies preferentially host
slower decline rate (brighter) SNe.  Riess et al.  (1999) also find that
SNe\,Ia with fast decline rates occur further from the host
galaxy centre.  Hamuy et al. (2000) present results from a larger
combined SNe sample, and find that brighter SNe occur in bluer stellar
environments (also noted by Branch et al.  1996), as well as some
evidence to suggest that the brightest SNe occur in the least luminous
galaxies. We note that no claim has yet been made that such correlations
would apply in such a manner as to reduce the significance of a
non-zero cosmological constant (Hamuy et al. 2000). 

Here, we extend these studies to include similar comparisons between
the host galaxies and properties of high-redshift SNe, where
environmental effects on the determination of the cosmological
parameters can be more easily explored. We present the preliminary
results of our study probing the environments of high-redshift SNe\,Ia
discovered via the SCP, and compare these high-redshift SNe/host galaxy
correlations with the local sample of SNe of Hamuy et al.  (1996a), the
same sample previously used in P99 for the determination of
cosmological parameters. Though larger local SNe\,Ia samples have
recently become available (Riess et al.  1999), in this preliminary
analysis we restrict ourselves to the above local sample for ease of
comparison with the P99 cosmological results, deferring a full analysis
to a later article (Sullivan et al., in prep.). 

\section{New Data and Host Galaxy Classification}

\subsection{\textit{HST} imaging}

\begin{figure}
\centering
\epsfig{file=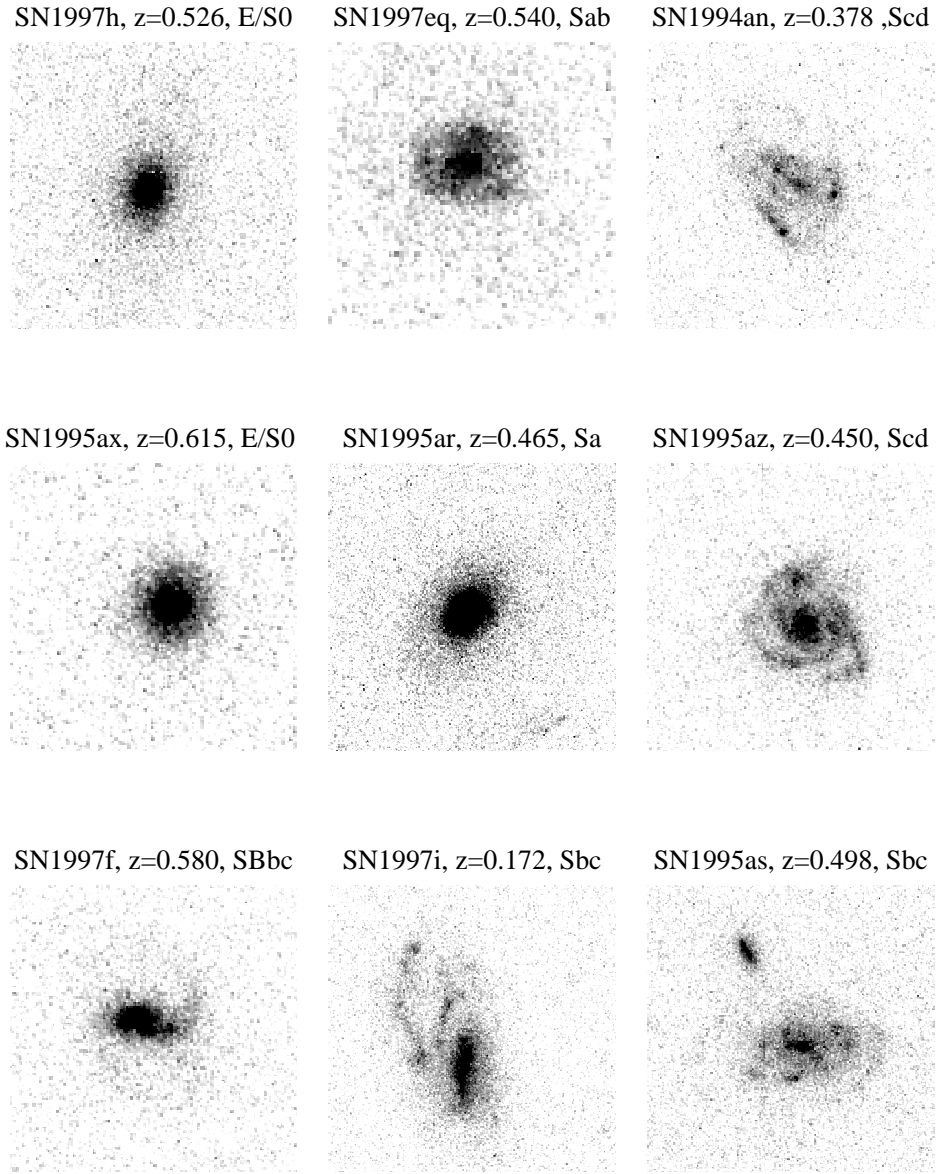,width=5in}
\caption{
  A selection of the SNe\,Ia host galaxies imaged using STIS on
  the \textit{HST}, together with the SNe redshift and host
  galaxy classification.}
\end{figure}

The primary component of our survey is \textit{HST} snapshot images of
the host galaxies of 54 SNe\,Ia discovered via the SCP, including those
42 SNe\,Ia used in P99. Each galaxy is being imaged using the STIS in
its 50CCD (clear aperture) mode, which approximates a broad $V+R+I$
bandpass. To date, 31 of the galaxies have been imaged, though in one
case the target field was missed by \textit{HST}. Each galaxy is imaged
in a $3\,\times\,434\rm{s}$ dithered series, and the data reduced using
the standard STScI pipeline.  Cosmic ray rejection and image
combination was performed using the \textit{Dither II} and
\textit{Drizzle} packages in \textsc{iraf} (Fruchter et al.  1997;
Fruchter \& Hook 1999). Of the 30 successfully targeted galaxies, 27
have an adequate S/N for visual classification within a broad
E/S0:Sab:Scd/Irr scheme (see Figure~1 for examples), 20 of which were
used in the Hubble diagram published by P99.

To locate the position and projected distance of the now-faded SNe from
the host galaxy centre, the \textit{HST} snapshot images were
cross-correlated with existing deep ground-based images from the
original discovery campaigns. Most of the STIS fields typically
contained 4-8 objects in common with the ground-based images, which
could be used to pinpoint the exact SN location on the STIS image
(typically to $0.3-0.4\,''$) for both a visual inspection of the SN
environment, and for subsequent spectroscopic campaigns.

\subsection{Keck-II ESI spectroscopic campaign}

\begin{figure}
\begin{center}
\parbox{2in}{
    \epsfig{file=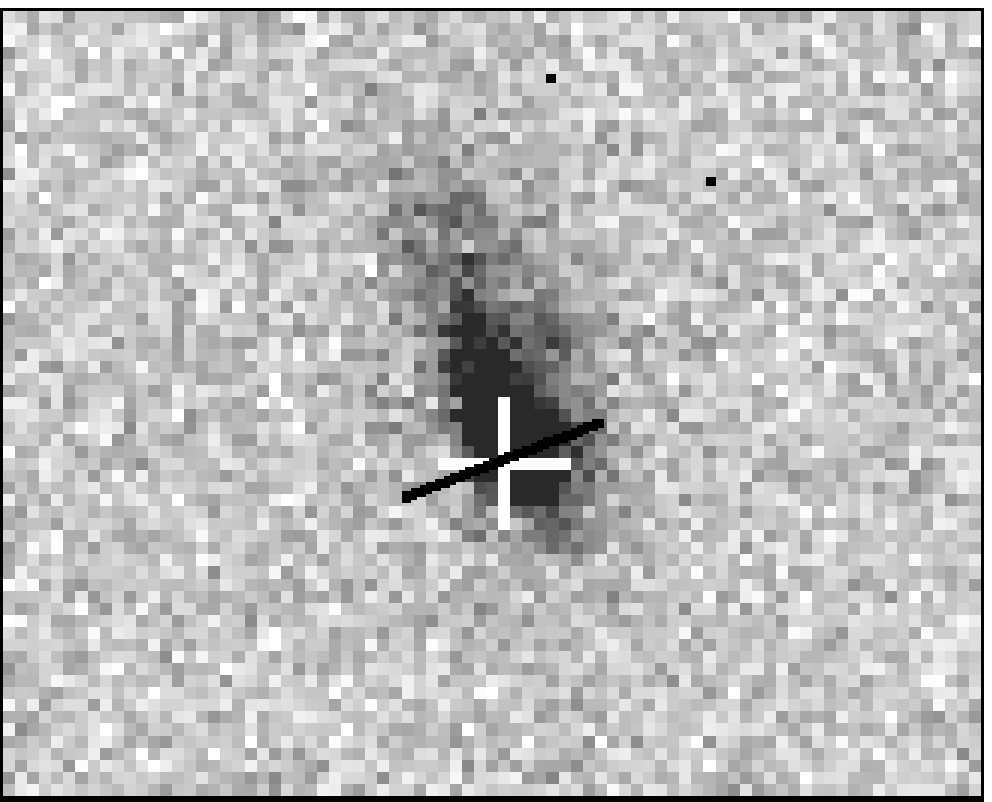,width=2in}}\hspace{0.1in}
\parbox{3in}{\epsfig{file=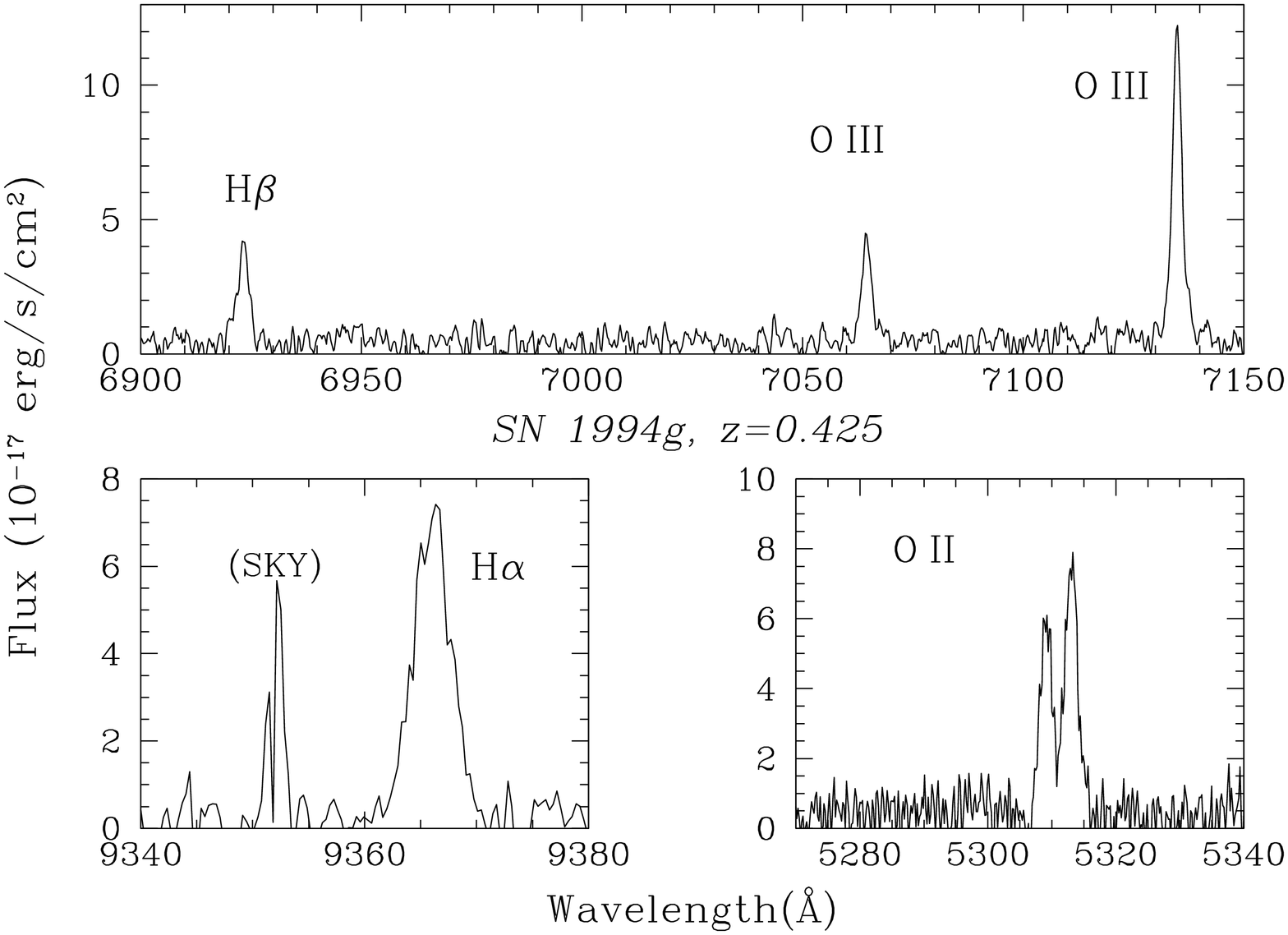,angle=270,width=3in}}
\parbox{2in}{ \epsfig{file=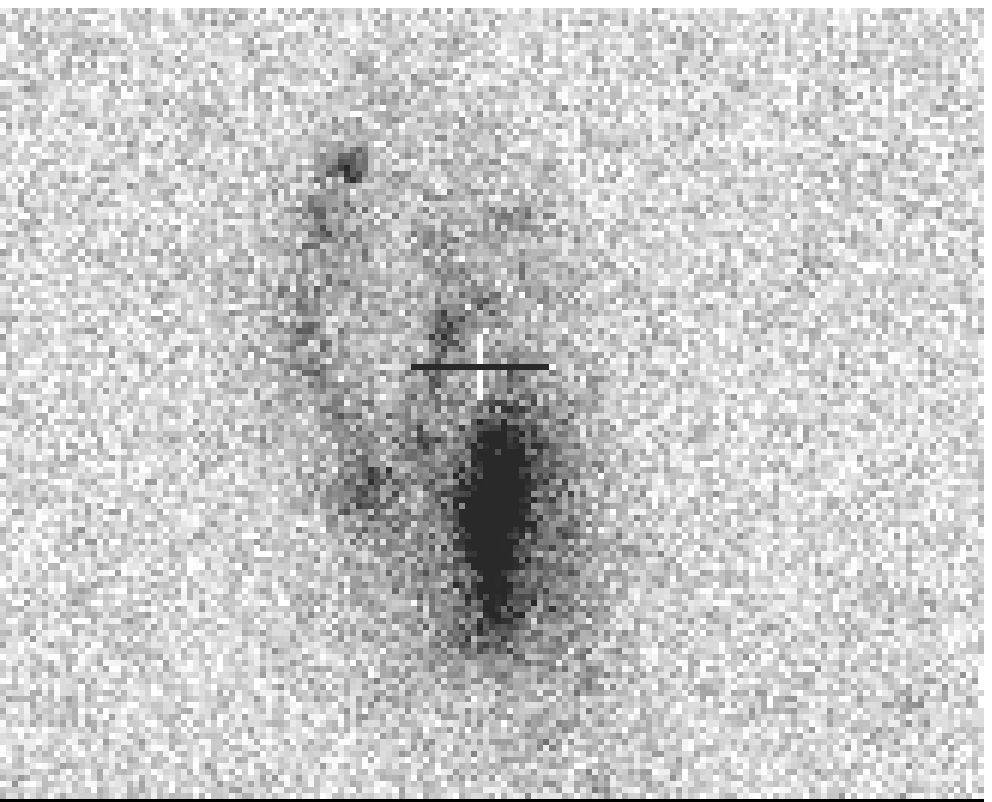,width=2in}}\hspace{0.1in}
\parbox{3in}{\epsfig{file=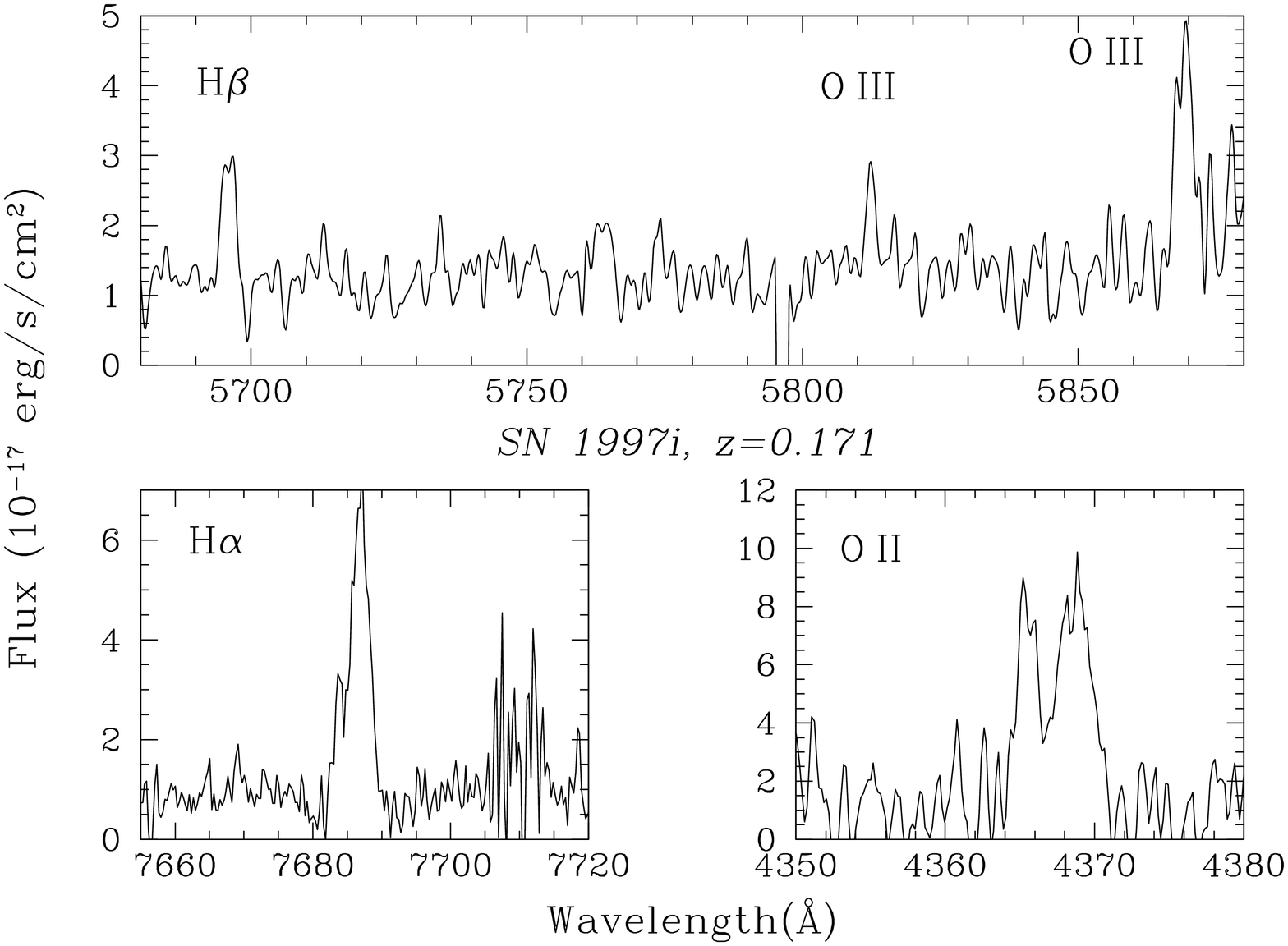,angle=270,width=3in}}

\caption{
  Two of the six host galaxies observed using ESI.  TOP:
  \textit{SN1994g}, $z=0.425$, 3600s integration. BOTTOM:
  \textit{SN1997i}, $z=0.171$, 1800s integration. These and the
  other spectra have enabled reddening estimates in 4 out of the 5 
  disk-based galaxies targeted.}
\label{esi}
\end{center}
\end{figure}

The existing host galaxy spectra were taken during the period when SN
light dominated that of the galaxy, making any studies of host galaxy
properties very difficult.  Using a subsample of our STIS-imaged
galaxies, we have therefore begun a second round of optical
spectroscopy using the Echellette Spectrograph Imager (ESI) on the
Keck-II telescope. The high spatial resolution of the STIS images and
the astrometric location of the SN position enables us to accurately
align the $20\,''\times1\,''$ ESI slit through the appropriate
projected region of the galaxy. To date, 6 host galaxies have been
studied in this way, of which 5 were Sc/d or irregular systems;
example spectra are shown in Figure~2.

The spectra allow an accurate spectral type (and confirmatory
redshift) to be assigned to each galaxy, and estimates of the spatial
variation in the star-formation characteristics and dust reddening
(via the H$\alpha$/H$\beta$ ratio) at different locations in the
vicinity of the SN. Spectra were characterised according to local
templates (c.f. Heyl et al. 1997). Whilst the spectroscopic sample is
currently small, when enlarged it will permit a further
characterisation of the galaxy population and the uncertainties in the
adopted reddening estimates (Sullivan et al., in prep).

\subsection{Galaxy Classification}

We assign a host galaxy type in 3 broad categories: spheroidal (E/S0),
spiral (Sa through Sc) and late (Scd/Irr), based on the synergy between
three diagnostics available to us:

\begin{enumerate}
\item[$\bullet$] The \textit{HST}-STIS morphology (20/42 SNe host 
     galaxies from P99 have this data, Figure~1),
\item[$\bullet$] The ESI or, where useful, pre-existing 
    spectroscopic data (6/42 SNe host galaxies from P99 have ESI 
    spectra, Figure~2),
\item[$\bullet$] The $R-I$ colour from the original reference
  SNe search images, i.e. those unaffected by SNe  (35/42 SNe
  host galaxies from P99 have adequate data).
\end{enumerate}

\noindent
Our primary diagnostics are the STIS imaging and ESI spectra; in all
available cases these agree. However, in cases where the STIS
morphology may be ambiguous (and no ESI spectrum is available), we
demand at least one of the other two diagnostics -- either the colour
or the pre-existing spectral type -- to agree with the STIS
classification. In total, 24/42 (57\%) P99 high-redshift SNe\,Ia host
galaxies have been classified according to this scheme. Useful statistics
of the combined low-z and high-z sample are given in Table~1.

\begin{center}
\begin{table}
\caption{
{\em Combined low and high-$z$ host galaxy classifications}. 
Low-$z$ classifications are taken from Hamuy et al. (2000). Values 
in parenthesis indicate the number of SNe used in the best-fit 
(`fit C') of P99. Also shown are the dispersions
of the residuals derived from the best-fit cosmology of P99 for the 
combined sample and the value of $\Lambda$ calculated by fitting SNe in each galaxy class.
The final two columns show the same but without stretch corrections applied to the SNe peak luminosity.}

\begin{center}
\begin{tabular}{cccccccc}
\tableline
Type & low-$z$ & high-$z$ & Total & Dispersion& Best $\Lambda$ & Dispersion & Best $\Lambda$\\
& & & & \multicolumn{2}{c}{(with stretch)} & \multicolumn{2}{c}{(no stretch)}\\
\tableline
Spheroidal & 7 (5) & 7 (7) & 14 (12) & 0.195 & 0.58 & 0.210 & 0.63 \\
Spiral & 8 (8) & 5 (4) & 13 (12) & 0.270 & 0.30 & 0.280 & 0.25 \\
Late/Irr & 3 (3) & 12 (11) & 15 (14) & 0.300 & 0.83 & 0.286 & 0.75 \\
\tableline
\tableline
\end{tabular}
\end{center}
\end{table}
\end{center}

\section{Current Results}

Here we present our preliminary findings and results for the first 24
SNe host galaxies classified as discussed above. As the dataset
continues to grow, the quantitative numerical results must be
considered provisional. Moreover, only half of the high-$z$ SNe hosts
used in the P99 analysis have so far been classified.

We first examine the Hubble diagram of P99 (their fig.~2), labelling each
SN according to its host galaxy type (Figure~3). The plot suggests that 
SNe occurring in spheroidal galaxies show less scatter than those in 
early-type spirals and late-type spirals or irregulars.

\begin{figure}
\centering
\epsfig{file=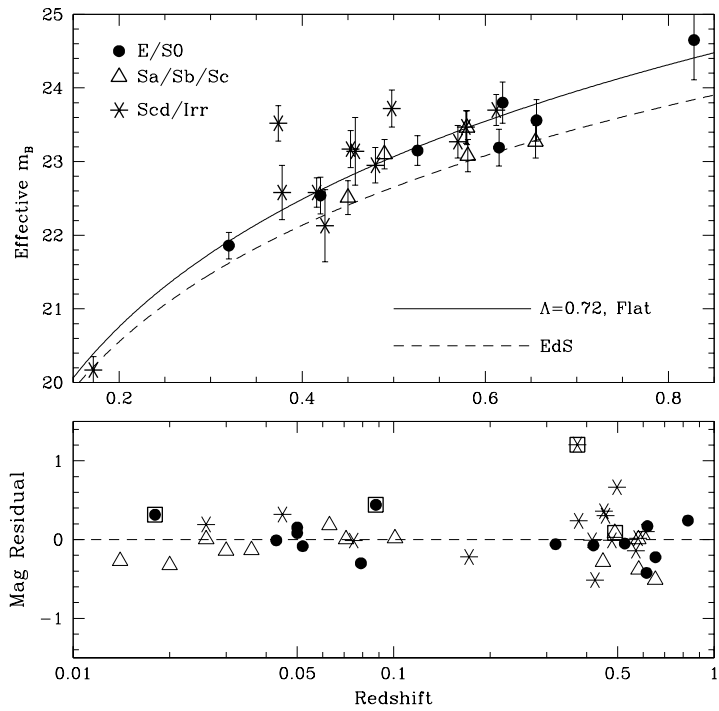,width=4in}
\label{fig_hubble}
\caption{
  The Hubble diagram of P99 plotted with each SN labelled according to the
  host galaxy type. The top panel shows the high-$z$ SNe only,
  while the lower panel plots the residuals from the adopted cosmology 
  (`fit C' of P99) for both high and low-$z$ SNe. Boxed points show 
  SNe excluded from the P99 solution.}
\end{figure}

This effect is quantified in Table~1, which shows for each galaxy
class the mean residual from the best-fit (`fit C') cosmology of P99
($\Lambda=0.72$, $\Omega=0.28$). We also estimate the implied change
in the inferred $\Lambda$ when SNe originating in each class are used
to establish the cosmological model (assuming a flat universe:
$\Lambda+\Omega=1$). For each host galaxy class in turn, a
$\Lambda$-dominated cosmology is preferred, though clearly the
confidence level is considerably reduced, particularly for SNe in
early-type spirals where the sample is smallest. Most importantly
however, SNe found in late-type galaxies possess the largest scatter
and show a slightly larger $\Lambda$ (i.e. the SNe are on average
fainter), as expected if residual dust extinction affected their
measures more than in the case of spheroidal galaxies.

We also investigated the Hubble diagram with each SN plotted according
to its projected distance from the host galaxy centre, but found no
significant trend. Best-fit cosmologies for SNe located at both small
projected distances show slightly smaller derived $\Lambda$-values,
but the solutions are all within $1\sigma$ of the P99 result.

\begin{figure}
\centering
\epsfig{file=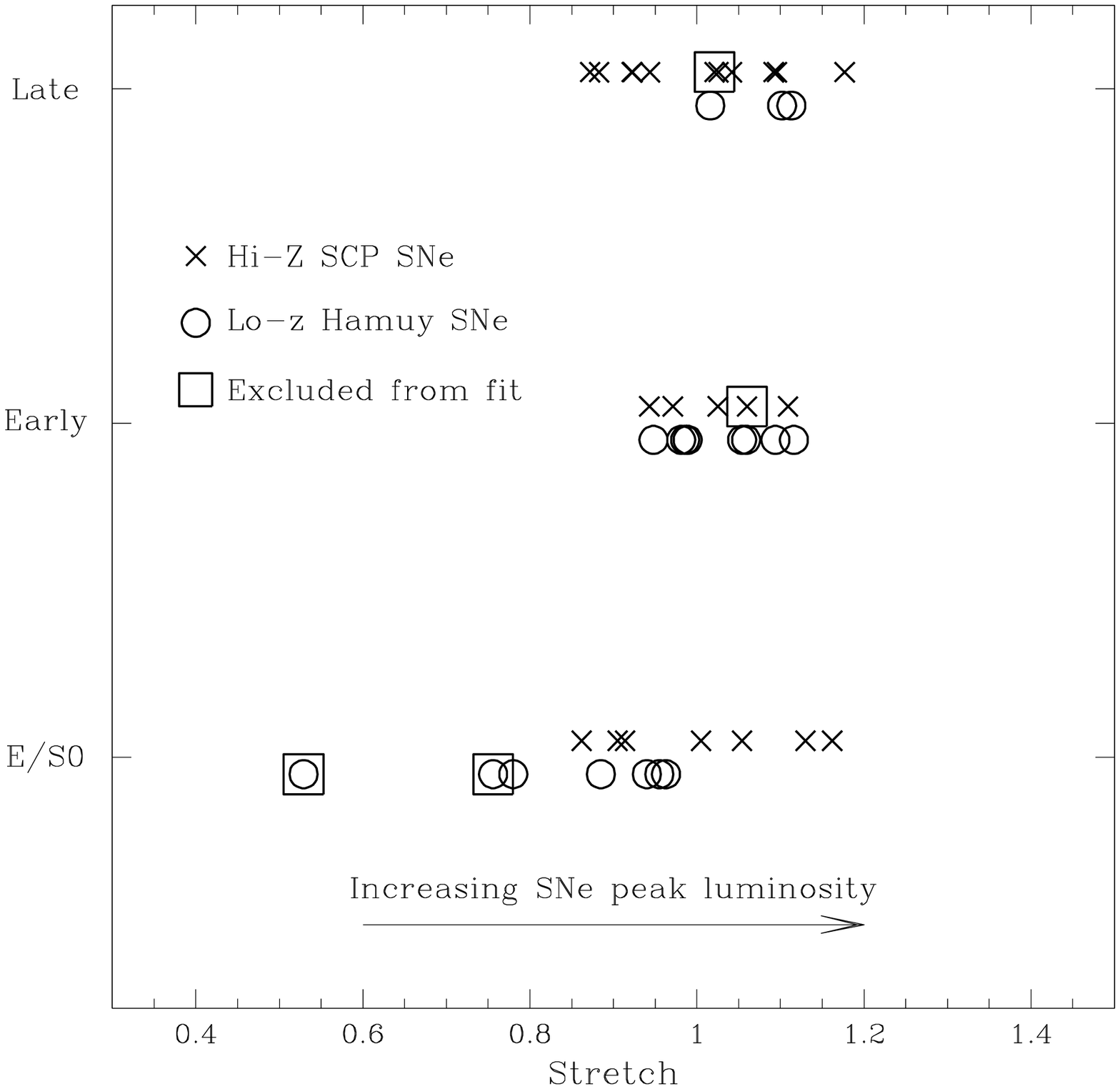,width=3in}
\caption{
  The distribution of SNe stretch with host galaxy type. The boxed
  points refer to SNe excluded from the primary fit of P99 either
  because they are stretch or residual outliers -- see P99 for further
  details. The low-$z$ SNe are taken from the Hamuy (1996a) sample.}
\end{figure}

There is evidence in local samples of SNe\,Ia of a correlation between
host galaxy type and SNe stretch correction (Hamuy et al. 2000), where
fainter (smaller stretch SNe) occur in E/S0 early-type galaxies (see
also Hamuy et al.  1996a, 2000; Riess et al. 1999).  In order to
ensure that the effect we see in Figure~3 is not simply related to the
use of inappropriate type-dependent stretch corrections, we repeat the
analysis but without making the stretch corrections to the SNe peak
luminosity (Table~1). We see that the quantitative result is
unchanged; the SNe in late-type galaxies still posses the most scatter
about the best-fit and show a slightly large value for $\Lambda$ than
those in E/S0 galaxies.

To investigate any stretch-host galaxy trends further, we plot in
Figure~4 the SN stretch as a function of morphological type for both
the low and high-$z$ P99 sample. As shown in P99, the stretch
distribution at high-redshift is narrower than that at low redshift,
but even within the narrow stretch range available at high-redshift,
we cannot clearly see any trends such as that which would be expected
in the spheroidal galaxies (cf Figure 3). At first sight, therefore,
Figure~4 suggests that SNe found at low and high redshift may be drawn
from two different populations (i.e. at high-redshift we see no trend
in stretch with galaxy type, whereas at low-redshift SNe with larger
stretches (i.e. brighter SNe) are found in later-type galaxies).
However, this result may not be that surprising. If we hypothesise
that intrinsically fainter SNe are found in older (E/S0) stellar
environments, then as we look to higher-redshift (i.e.  viewing
galaxies at an earlier stage in their evolution) we would not expect
to see the fainter SNe found in local, older, E/S0 galaxies (see Hamuy
et al.  2000).

Figure~5 shows the SN stretch plotted as a function of projected
distance from the host galaxy. Here a weak trend is found suggesting
that SNe at greater projected distances may be fainter, similar to
that seen in the local sample of Riess et al. (1999).  This could
possibly arise if, within a distribution of luminosities, fainter SNe
were systematically missed close to the galaxy core.

\section{Conclusions}

The preliminary results of our investigation continue to support the
conclusions of P99. However, the Hubble diagram categorised by galaxy
class offers new insight into the origin of the scatter in the
diagram. We find the scatter about the best-fit cosmology is greater
when determined from SNe which occurred in late-type/irregular
galaxies. Minimal scatter is found for SNe drawn from E/S0 galaxies as
expected if dust is largely absent in these systems. A non-zero
$\Lambda$ is supported by SNe arising from each galaxy type
individually, with late-type galaxies implying a larger $\Lambda$ than
E/S0 galaxies as might be expected when simple dust models are
considered. Finally, little difference is found when the Hubble
diagram is categorised according to SNe located at large and small
projected distances from their host galaxy.

\begin{figure}
\centering
\epsfig{file=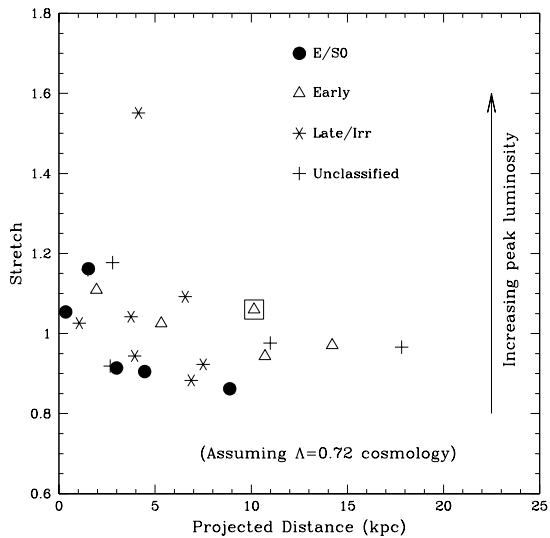,width=3in}
\caption{
  The distribution of SNe stretch with projected distance from host
  galaxy centre. Only galaxies with a \textit{HST} image are shown.
  The boxed points refer to SNe excluded from the primary fit of P99
  either because they are stretch or residual outliers. We see a
  slight trend with SNe located at a greater projected distance having
  a dimmer peak luminosity, similar to fig.~5c in Riess et al.
  (1999).}
\label{fig_stretch_distance}
\end{figure}

In the future, we will attempt to further subdivide our emission line
galaxies into two further types according to the spectral
characteristics. By using the high resolution of the ESI, we can
classify emission line galaxies according to the spatial variance of
their H$\alpha$/H$\beta$ ratio, describing how the dust extinction in
such objects varies according to the location inside the galaxy. We
will attempt to generate a subsample of these objects which posses a
dust distribution that shows large variances and examine whether these
objects are responsible for the significant scatter in the current
Hubble plot (Figure~3) among the late-type emission line galaxies. Those
with smoother dust distributions might be expected to have less
intrinsic scatter in the SNe peak luminosities (due to uncertain dust
corrections), typical of those SNe arising from E/S0 galaxies.

While these initial results are promising, the current sample size is
only around half that used in the full Hubble diagram in P99, and it
is clear that further \textit{HST} imaging and spectroscopic studies of
the host galaxy population are needed to confirm the initial findings
presented here.

\acknowledgements We thank Saul Perlmutter, Peter Nugent, Piero Madau
and the Supernovae Cosmology Project team for the many useful
discussions and support offered for this work.

\end{document}